\documentclass{webofc}
\usepackage[varg]{txfonts}   
\usepackage{natbib}
\def\aj{AJ}%
\def\apj{ApJ}%
\def\aap{A\&A}%
\def\mnras{MNRAS}%
\def\pasj{PASJ}%
\def\ssr{Space~Sci.~Rev.}%
\begin{document}
\title{NIKA2 observations of 3 low-mass galaxy clusters at $z \sim 1$: pressure profile and $Y_{\rm SZ} - M$ relation}
          
\author{
         \lastname{R.~Adam}\inst{\ref{OCA},\ref{LLR}}\fnsep\thanks{\email{remi.adam@oca.eu}}
\and  \lastname{M.~Ricci}\inst{\ref{APC},\ref{LAPP},\ref{OCA}}
\and  \lastname{D.~Eckert}\inst{\ref{geneva}}
\and  \lastname{P.~Ade}\inst{\ref{Cardiff}}
\and  \lastname{H.~Ajeddig}\inst{\ref{CEA}}
\and  \lastname{B.~Altieri}\inst{\ref{ESAC}}
\and  \lastname{P.~Andr\'e}\inst{\ref{CEA}}
\and  \lastname{E.~Artis}\inst{\ref{LPSC}}
\and  \lastname{H.~Aussel}\inst{\ref{CEA}}
\and  \lastname{A.~Beelen}\inst{\ref{LAM}}
\and  \lastname{C.~Benoist}\inst{\ref{OCA}}
\and  \lastname{A.~Beno\^it}\inst{\ref{Neel}}
\and  \lastname{S.~Berta}\inst{\ref{IRAMF}}
\and  \lastname{L.~Bing}\inst{\ref{LAM}}
\and  \lastname{M.~Birkinshaw}\inst{\ref{Bristol}}\thanks{Deceased}
\and  \lastname{O.~Bourrion}\inst{\ref{LPSC}}
\and  \lastname{D.~Boutigny}\inst{\ref{LAPP}}
\and  \lastname{M.~Bremer}\inst{\ref{Bristol}}
\and  \lastname{M.~Calvo}\inst{\ref{Neel}}
\and  \lastname{A.~Cappi}\inst{\ref{OCA},\ref{BolognaObs}}
\and  \lastname{A.~Catalano}\inst{\ref{LPSC}}
\and  \lastname{M.~De~Petris}\inst{\ref{Roma}}
\and  \lastname{F.-X.~D\'esert}\inst{\ref{IPAG}}
\and  \lastname{S.~Doyle}\inst{\ref{Cardiff}}
\and  \lastname{E.~F.~C.~Driessen}\inst{\ref{IRAMF}}
\and  \lastname{L.~Faccioli}\inst{\ref{CEA}}
\and  \lastname{C.~Ferrari}\inst{\ref{OCA}}
\and  \lastname{F.~Gastaldello}\inst{\ref{Milan}}
\and  \lastname{P.~Giles}\inst{\ref{Sussex}}
\and  \lastname{A.~Gomez}\inst{\ref{CAB}}
\and  \lastname{J.~Goupy}\inst{\ref{Neel}}
\and  \lastname{O.~Hahn}\inst{\ref{OCA}}
\and  \lastname{C.~Hanser}\inst{\ref{LPSC}}
\and  \lastname{C.~Horellou}\inst{\ref{Onsala}}
\and  \lastname{F.~K\'eruzor\'e}\inst{\ref{Argonne}}
\and  \lastname{E.~Koulouridis}\inst{\ref{Athene},\ref{CEA}}
\and  \lastname{C.~Kramer}\inst{\ref{IRAMF}}
\and  \lastname{B.~Ladjelate}\inst{\ref{IRAME}}
\and  \lastname{G.~Lagache}\inst{\ref{LAM}}
\and  \lastname{S.~Leclercq}\inst{\ref{IRAMF}}
\and  \lastname{J.-F.~Lestrade}\inst{\ref{LERMA}}
\and  \lastname{J.F.~Mac\'ias-P\'erez}\inst{\ref{LPSC}}
\and  \lastname{S.~Madden}\inst{\ref{CEA}}
\and  \lastname{B.~Maughan}\inst{\ref{Bristol}}
\and  \lastname{S.~Maurogordato}\inst{\ref{OCA}}
\and  \lastname{A.~Maury}\inst{\ref{CEA}}
\and  \lastname{P.~Mauskopf}\inst{\ref{Cardiff},\ref{Arizona}}
\and  \lastname{A.~Monfardini}\inst{\ref{Neel}}
\and  \lastname{M.~Mu\~noz-Echeverr\'ia}\inst{\ref{LPSC}}
\and  \lastname{F.~Pacaud}\inst{\ref{Bonn}}
\and  \lastname{L.~Perotto}\inst{\ref{LPSC}}
\and  \lastname{M.~Pierre}\inst{\ref{CEA}}
\and  \lastname{G.~Pisano}\inst{\ref{Roma}}
\and  \lastname{E.~Pompei}\inst{\ref{ESO}}
\and  \lastname{N.~Ponthieu}\inst{\ref{IPAG}}
\and  \lastname{V.~Rev\'eret}\inst{\ref{CEA}}
\and  \lastname{A.~Rigby}\inst{\ref{Cardiff}}
\and  \lastname{A.~Ritacco}\inst{\ref{INAF},\ref{ENS}}
\and  \lastname{C.~Romero}\inst{\ref{UPenn}}
\and  \lastname{H.~Roussel}\inst{\ref{IAP}}
\and  \lastname{F.~Ruppin}\inst{\ref{IP2I}}
\and  \lastname{M.~Sereno}\inst{\ref{BolognaObs},\ref{BolognaDepP}}
\and  \lastname{K.~Schuster}\inst{\ref{IRAMF}}
\and  \lastname{A.~Sievers}\inst{\ref{IRAME}}
\and  \lastname{G.~Tintor\'e Vidal}\inst{\ref{LLR}}
\and  \lastname{C.~Tucker}\inst{\ref{Cardiff}}
\and  \lastname{R.~Zylka}\inst{\ref{IRAMF}}
}

\institute{
Laboratoire Lagrange, Universit\'e C\^ote d'Azur, Observatoire de la C\^ote d'Azur, CNRS, Blvd de l'Observatoire, CS 34229, 06304 Nice cedex 4, France
  \label{OCA}
  \and
 LLR, CNRS, \'Ecole Polytechnique, Institut Polytechnique de Paris
  \label{LLR}
  \and
  Universit\'e Paris Cit\'e, CNRS, Astroparticule et Cosmologie, F-75013 Paris, France
  \label{APC}
  \and
Laboratoire d'Annecy de Physique des Particules, Universit\'e Savoie Mont Blanc, CNRS/IN2P3, F-74941 Annecy, France
  \label{LAPP}
  \and
  Department of Astronomy, University of Geneva, ch. d'Ecogia 16, CH-1290 Versoix, Switzerland
   \label{geneva}
   \and
    School of Physics and Astronomy, Cardiff University, Queen’s Buildings, The Parade, Cardiff CF24 3AA, UK
  \label{Cardiff}
  \and
Universit\'e Paris-Saclay, Universit\'e Paris Cit\'e, CEA, CNRS, AIM, F-91191, Gif-sur-Yvette, France (2022)
  \label{CEA}
  \and
European Space Astronomy Centre (ESA/ESAC), Operations Department, Villanueva de la Can\~{a}da, Madrid, Spain
\label{ESAC}
  \and
  Laboratoire de Physique Subatomique et de Cosmologie, Universit\'e Grenoble Alpes, CNRS/IN2P3, 53, avenue des Martyrs, Grenoble, France
  \label{LPSC}
\and
Aix Marseille Universit\'e, CNRS, LAM (Laboratoire d'Astrophysique de Marseille) UMR 7326, 13388, Marseille, France
  \label{LAM}
  \and
Institut N\'eel, CNRS and Universit\'e Grenoble Alpes, France
  \label{Neel}
    \and
Institut de RadioAstronomie Millim\'etrique (IRAM), Grenoble, France
  \label{IRAMF}
  \and
HH Wills Physics Laboratory, University of Bristol, Tyndall Avenue, Bristol, BS8 1TL, UK
\label{Bristol} 
\and
  Istituto Nazionale di Astrofisica (INAF) - Osservatorio di Astrofisica e Scienza dello Spazio (OAS), via Gobetti 93/3, I-40127 Bologna, Italy
\label{BolognaObs}
\and
Dipartimento di Fisica, Sapienza Universit\`a di Roma, Piazzale Aldo Moro 5, I-00185 Roma, Italy
  \label{Roma}
\and
Univ. Grenoble Alpes, CNRS, IPAG, F-38000 Grenoble, France 
  \label{IPAG}
 \and
INAF - IASF Milan, via A. Corti 12, I-20133 Milano,  Italy
\label{Milan}  
\and
Department of Physics and Astronomy, University of Sussex, Brighton BN1 9QH, UK
\label{Sussex} 
\and
Centro de Astrobiolog\'ia (CSIC-INTA), Torrej\'on de Ardoz, 28850 Madrid, Spain
\label{CAB}
\and
Department of Space, Earth and Environment, Chalmers University of Technology, Onsala Space Observatory, SE-439 92 Onsala,
Sweden
\label{Onsala}
\and
High Energy Physics Division, Argonne National Laboratory, 9700 South Cass Avenue, Lemont, IL 60439, USA
\label{Argonne}
\and
Institute for Astronomy \& Astrophysics, Space Applications \& Remote
Sensing, National Observatory of Athens, GR-15236 Palaia Penteli,Greece
  \label{Athene}
  \and
Institut de RadioAstronomie Millim\'etrique (IRAM), Granada, Spain 
\label{IRAME}
\and 
LERMA, Observatoire de Paris, PSL Research University, CNRS, Sorbonne Universités, UPMC Univ. Paris 06, 75014 Paris, France
  \label{LERMA}  
\and
School of Earth and Space Exploration and Department of Physics, Arizona State University, Tempe, AZ 85287
  \label{Arizona}
\and 
Argelander Institut f\"ur Astronomie, Universit\"at Bonn, Auf dem Huegel 71, DE-53121 Bonn, Germany
\label{Bonn}
\and
European Southern Observatory, Alonso de Cordova 3107, Vitacura,
19001 Casilla, Santiago 19, Chile
\label{ESO}
\and
INAF-Osservatorio Astronomico di Cagliari, Via della Scienza 5, 09047 Selargius, Italy
\label{INAF}
\and
LPENS, Ecole Normale Sup\'erieure, 24 rue Lhomond, 75005, Paris (FR) 
\label{ENS}
\and
Department of Physics and Astronomy, University of Pennsylvania, 209 South 33rd Street, Philadelphia, PA 19104, USA
\label{UPenn}
\and 
Institut d'Astrophysique de Paris, Sorbonne Universit\'es, UPMC Univ. Paris 06, CNRS UMR 7095, 75014 Paris, France 
  \label{IAP}
\and
Univ. Lyon, Univ. Claude Bernard Lyon 1, CNRS/IN2P3, IP2I Lyon, 69622 Villeurbanne, France
\label{IP2I}
\and
INFN, Sezione di Bologna, viale Berti Pichat 6/2, 40127 Bologna, Italy
\label{BolognaDepP}
\and
Institut d'Astrophysique Spatiale (IAS), CNRS and Universit\'e Paris Sud, Orsay, France
  \label{IAS}
}

\abstract{Three galaxy clusters selected from the XXL X-ray survey at high redshift and low mass ($z\sim1$ and $M_{500} \sim 1-2 \times 10^{14}$ M$_{\odot}$) were observed with NIKA2 to image their Sunyaev-Zel'dovich effect (SZ) signal. They all present an SZ morphology, together with the comparison with X-ray and optical data, that indicates dynamical activity related to merging events. Despite their disturbed intracluster medium, their high redshifts, and their low masses, the three clusters follow remarkably well the pressure profile and the SZ flux-mass relation expected from standard evolution. This suggests that the physics that drives cluster formation is already in place at $z \sim 1$ down to $M_{500} \sim 10^{14}$ M$_{\odot}$.}

\maketitle

\section{Introduction}
Galaxy clusters are important astrophysical objects that provide insights into the formation and evolution of cosmic structures. Understanding the properties of their intracluster medium (ICM) is crucial for both cosmology and astrophysics. The thermal pressure profile of the ICM provides valuable information about the matter distribution within galaxy clusters. It reflects how the gas is compressed within the cluster's gravitational potential well. Additionally, the $Y_{\rm SZ} - M$ scaling relation is an important tool for measuring cluster masses, as it relates to the Sunyaev-Zel'dovich \citep[SZ, see][]{Sunyaev1972} flux ($Y_{\rm SZ}$, which directly measures the thermal energy of the ICM) to the total cluster mass, $M$ \citep[see][for a review]{Mroczkowski2019}.

While the thermal pressure profile and the $Y_{\rm SZ} - M$ scaling relation have been extensively studied up to intermediate redshifts \citep[e.g.,][]{Arnaud2010,PlanckV2013}, there is a lack of detailed observations in the low-mass, high-redshift regime. In this proceeding, we bridge this observational gap by presenting detailed SZ observations of three low-mass ($M_{500} \sim 2 \times 10^{14}$ M$_{\odot}$) galaxy clusters at $z\sim1$ selected from the XXL X-ray survey. These clusters, namely XLSSC 072, XLSSC 100, and XLSSC 102 were imaged using the NIKA2 millimeter camera on the IRAM 30-meter telescope at 150 GHz and 260 GHz \citep{Adam2018a}.

By combining the SZ observations with X-ray and optical data, we investigate the dynamic state of the clusters and derive their thermal pressure profiles. We also extract the SZ fluxes and compare them with expectations based on the standard evolution of the ICM properties calibrated on nearby massive systems. This analysis allows us to explore any deviations from the expected evolution in the ICM properties for low-mass clusters at high redshifts, where astrophysical processes are expected to have a more significant impact \citep{Fakhouri2010,Pop2022}.

This proceeding is organized as follows: Section~\ref{sec:target_and_obs} presents the cluster selection and the observations, Section~\ref{sec:Analysis} discusses the data analysis used to address the cluster dynamical state, extract the pressure profile and the $Y_{\rm SZ}-M$ relation, in Section \ref{sec:conclusions}, we summarize our main findings and conclusions. See \cite{adam2023xxl} for the full results. See also \citep{Ricci2020} for the detailed analysis of the first cluster XLSSC~102.

\section{Target selection and observations}\label{sec:target_and_obs}
We selected our target clusters from the XXL survey \citep{Pierre2016}. The XXL survey's selection function enabled the identification of clusters at low mass and high redshift \citep{Pacaud2016}. We specifically focused on the securely detected XXL clusters (C1) from the northern region (XXL-N), observable from the IRAM 30m telescope. Optical detections using galaxy overdensity were used for independent confirmation, with robust spectroscopic redshift estimates \citep{Adami2018}. We selected three clusters, XLSSC 072, XLSSC 100, and XLSSC 102, at a redshift $z \sim 1$ and $M_{500} \sim 1-2 \times 10^{14}$ M$_{\odot}$, for NIKA2 observations. 

The observations were conducted between January 2018 and February 2020. XLSSC 072, XLSSC 100, and XLSSC 102 were observed for 10, 10, and 6.6 hours, respectively. The data reduction followed the procedure described in \cite{Adam2015} and \cite{Adam2016}. The data analysis included estimating the astrophysical signal filtering induced by the data reduction and deriving the noise statistical properties through power spectrum analysis and Monte Carlo simulations. Overall, the observational details and data reduction process for all three clusters were consistent with the characteristics of the NIKA2 instrument \citep{Perotto2020}. 

After smoothing the NIKA2 data to an effective resolution of 27 arcsec, XLSSC~072, XLSSC~100, and XLSSC~102, are detected with a peak signal-to-noise ratio (S/N) of -9.7, -9.2, and -6.9, respectively. Radio and submillimeter point sources were removed from the SZ images prior to further analysis.

\section{Analysis and results}\label{sec:Analysis}
\subsection{Dynamical state}
The SZ data were combined with X-ray (XXL survey \cite{Pierre2016}) and optical images (CFHTLS \cite{Gwyn2012} and HSC \cite{Aihara2022}) and analyzed to trace the ICM thermal pressure, the ICM thermal density, and the collisionless galaxy population, respectively. Comparing these tracers provides valuable insights into the clusters' dynamical properties.

The multiwavelength analysis shows that all three clusters exhibit deviations from spherical symmetry, indicating the presence of disturbances in the gas and galaxy distribution, likely due to merging events. The SZ and X-ray signals agree well on large scales but may differ on smaller scales, suggesting local compressions caused by merging. The surface brightness distributions of both SZ and X-ray signals are relatively flat, indicating dynamically disturbed systems without prominent X-ray peaks associated with relaxed clusters.

\subsection{Mass measurement}
The masses were estimated using two main methods.
\begin{enumerate}
\item Combining the X-ray derived density profile \citep{Eckert2011,Eckert2020} and the SZ-derived pressure profile (see below) via the hydrostatic equilibrium assumption,
\begin{equation}
M_{\rm HSE}(r) = -\frac{r^2}{\mu_{\rm gas} m_{\rm p} n_{\rm e}(r) G} \frac{d P_e(r)}{dr}.
\label{eq:MHSE}
\end{equation}
This provides a direct measurement but is more sensitive to systematics (hydrostatic mass bias, clumping, etc).
\item Using the scaling relation between $Y_{\rm X,500} = M_{\rm gas}(R_{500}) \ k_{\rm B}T$ and the mass from \cite{Arnaud2010},
\begin{equation}
E(z)^{-2/3} \left(\frac{Y_{{\rm X}, 500}}{2 \times 10^{14} {\rm M}_{\odot} {\rm keV}}\right) = 10^{0.376} \times \left(\frac{M_{{\rm HSE},500}}{6\times 10^{14} {\rm M}_{\odot}}\right)^{1.78}.
\label{eq:Yx-M_scaling}
\end{equation}
The gas mass $M_{\rm gas}$ and temperature $k_{\rm B}T$ were extracted from X-ray data. This provides a robust mass proxy but relies on local calibration.
\end{enumerate}

We obtain direct hydrostatic masses of $M_{\rm HSE,500} = 1.93_{-0.30}^{+0.38}$, $2.55_{-0.73}^{+1.46}$ and $1.12_{-0.20}^{+0.22}$ M$_{\odot}$ for XLSSC~072, XLSSC~100 and XLSSC~102, respectively. The masses based on the $Y_{\rm X}$ proxy are $M_{Y_{\rm X}, 500} = 1.98_{-0.17}^{+0.31}$, $2.13_{-0.33}^{+0.49}$, and $1.88_{-0.19}^{+0.28}$ M$_{\odot}$ for the same clusters. They agree well within error bars, although a $2 \sigma$ tension is observed for XLSSC~102 (see discussion hereafter).

\subsection{Pressure profile}
The pressure profile of the three clusters was measured using three different methodologies.
\begin{enumerate}
\item Forward modeling of the NIKA2 data using a gNFW pressure profile \citep{Nagai2007}, given by 
\begin{equation}
P_e(r) = \frac{P_0}{\left(\frac{r}{r_p}\right)^c \left(1 + \left(\frac{r}{r_p}\right)^a\right)^{\frac{b-c}{a}}}.
\label{eq:gNFW}
\end{equation}
\item Non-parametric fitting of a binned pressure profile. We defined the pressure at five radii, logarithmically spaced from 50 kpc to 1 Mpc.
\item Forward modeling of the NIKA2 data using an NFW \citep{Navarro1996} hydrostatic mass profile combined with the density profile derived from X-ray data. According to equation~\ref{eq:MHSE}, the pressure is modeled as
\begin{equation}
P_e(r) = P_e(r_0) + \int_r^{r_0} \frac{\mu_{\rm gas} m_{\rm p} G n_e(r^{\prime}) M_{\rm HSE}(r^{\prime})}{{r^{\prime}}^2}dr^{\prime},
\label{eq:NFW}
\end{equation}
\end{enumerate}

\begin{figure*}
	\centering
	\includegraphics[width=0.99\textwidth]{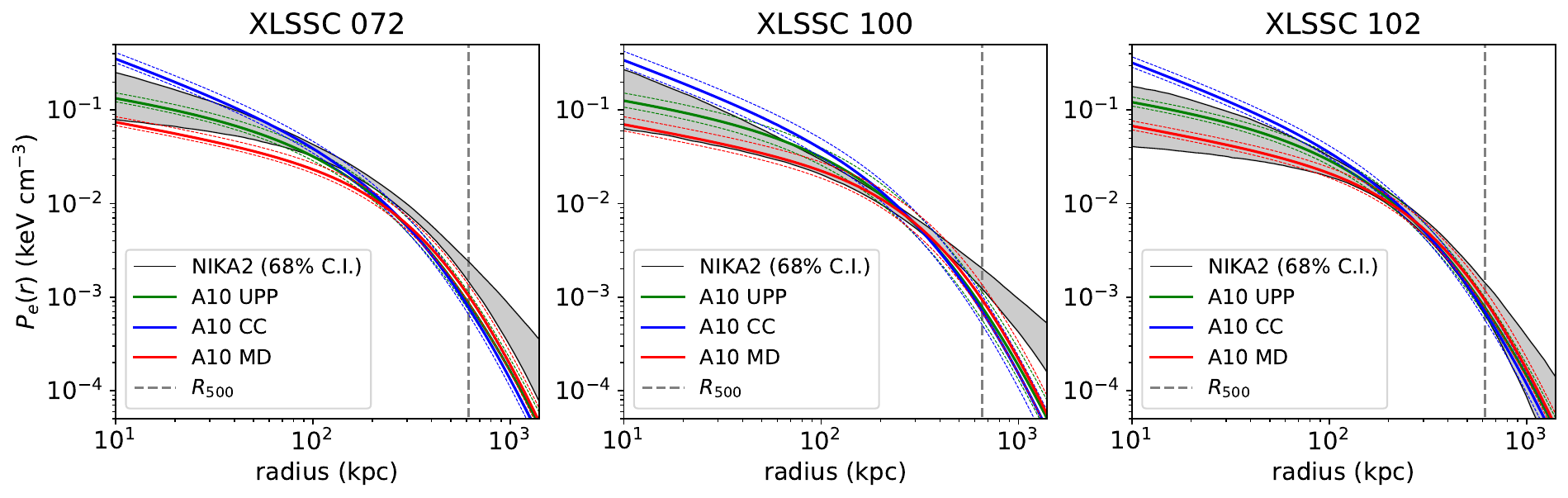}
	\caption{NIKA2 derived pressure profile and comparison with expectations from the universal pressure profile (UPP) from \cite{Arnaud2010}, as well as morphologically disturbed (MD) and cool-core systems (CC), given $Y_{\rm X}$-based masses. Figure extracted from \citep{adam2023xxl}.}
\label{figure}
\end{figure*}

Assuming standard evolution and given our mass estimates, the pressure profiles align well with the dynamical state analysis. Alternatively, considering the prior knowledge of the cluster dynamical states, the data agree favorably with the pressure profile calibrated on low-redshift clusters \citep{Arnaud2010} and scaled to low mass and high redshift using standard evolution (see Figure~\ref{figure}). All the pressure profiles recovered with the various methodologies exhibit excellent agreement within the uncertainties at all radii. Most of the uncertainty is due to the difficulty in having precise and robust mass measurements at these redshifts and mass scales.

\subsection{Scaling relation}
Our targets, XLSSC072, XLSSC100, and XLSSC~102, lie at the low-mass end of the Planck calibration sample \citep{PlanckXX2014} but have higher redshifts (around $z\sim1$) compared to the Planck clusters (around $z\sim0.2$). When using the $Y_{\rm X}-M$ relation for mass estimation, they follow the scaling relation remarkably well.
When direct hydrostatic equilibrium mass measurements are used, XLSSC102 deviates from the scaling relation by approximately $2\sigma$. However, this deviation may be attributed to systematic uncertainties in mass measurement due to the complex morphology and dynamical state of this cluster or a significant hydrostatic mass bias, as discussed in detail in \cite{Ricci2020}. The other clusters agree very well with the relation.

\section{Conclusions}\label{sec:conclusions}
Our study aimed to investigate the SZ structure and properties of the ICM in three low-mass galaxy clusters at a redshift $z \sim 1$. By utilizing the NIKA2 camera, we obtained valuable resolved SZ data for these clusters, representing some of the first detailed observations of such low-mass systems at high redshifts.

Our analysis revealed evidence of ongoing merging activity in all three clusters, as indicated by their disturbed morphologies and deviations from compact spherically symmetric distributions. The pressure profiles of these clusters, rescaled according to standard evolution in mass and redshift, were found to be in agreement with those of local dynamically disturbed systems. Despite their perturbed ICM, low masses, and high redshifts, we did not observe any significant deviations in the $Y_{\rm SZ}-M$ scaling relation for our target clusters. While the sample size is not sufficient to draw statistical conclusions, these findings provide an initial indication that the scaling relations are stable down to masses of $M_{500}\sim2\times10^{14}$ M$_{\odot}$ and redshifts of $z\sim1$. This may indicate that the cluster formation physics is already in place down to these masses and up to these redshifts

It is important to note that the comparison of the pressure profiles and scaling relation to those of local samples is limited by uncertainties in the mass estimates. This highlights the challenges in accurately determining and robustly estimating the mass in this relatively unexplored regime.


\begin{thebibliography}{22}

\bibitem{Sunyaev1972}
R.A. {Sunyaev}, Y.B. {Zeldovich}, Comments on Astrophysics and Space Physics \textbf{4}, 173 (1972)

\bibitem{Mroczkowski2019} 
T.~{Mroczkowski}, D.~{Nagai}, K.~{Basu}, J.~{Chluba}, J.~{Sayers}, R.~{Adam}, E.~{Churazov}, A.~{Crites}, L.~{Di Mascolo}, D.~{Eckert} et~al., \ssr\ \textbf{215}, 17 (2019)

\bibitem{Arnaud2010} 
M.~{Arnaud}, G.W. {Pratt}, R.~{Piffaretti}, H.~{B{\"o}hringer}, J.H. {Croston}, E.~{Pointecouteau}, \aap\ \textbf{517}, A92 (2010)

\bibitem{PlanckV2013} 
{Planck Collaboration}, P.A.R. {Ade}, N.~{Aghanim}, M.~{Arnaud}, M.~{Ashdown}, F.~{Atrio-Barandela}, J.~{Aumont}, C.~{Baccigalupi}, A.~{Balbi}, A.J. {Banday} et~al., \aap\ \textbf{550}, A131 (2013)

\bibitem{Adam2018a}
R.~{Adam}, A.~{Adane}, P.A.R. {Ade}, P.~{Andr{\'e}}, A.~{Andrianasolo}, H.~{Aussel}, A.~{Beelen}, A.~{Beno{\^\i}t}, A.~{Bideaud}, N.~{Billot} et~al., \aap\ \textbf{609}, A115 (2018)

\bibitem{Fakhouri2010}
O.~{Fakhouri}, C.P. {Ma}, M.~{Boylan-Kolchin}, \mnras\ \textbf{406}, 2267 (2010)

\bibitem{Pop2022}
A.R. {Pop}, L.~{Hernquist}, D.~{Nagai}, R.~{Kannan}, R.~{Weinberger}, V.~{Springel}, M.~{Vogelsberger}, D.~{Nelson}, R.~{Pakmor}, A.~{Pillepich} et~al., arXiv e-prints arXiv:2205.11528 (2022)

\bibitem{adam2023xxl}
R.~Adam, M.~Ricci, D.~Eckert, P.~Ade, H.~Ajeddig, B.~Altieri, P.~André, E.~Artis, H.~Aussel, A.~Beelen et~al., \aap\ (accepted), arXiv e-prints arXiv:2310.05819 (2023)

\bibitem{Ricci2020}
M.~{Ricci}, R.~{Adam}, D.~{Eckert}, P.~{Ade}, P.~{Andr{\'e}}, A.~{Andrianasolo}, B.~{Altieri}, H.~{Aussel}, A.~{Beelen}, C.~{Benoist} et~al., \aap\ \textbf{642}, A126 (2020)

\bibitem{Pierre2016}
M.~{Pierre}, F.~{Pacaud}, C.~{Adami}, S.~{Alis}, B.~{Altieri}, N.~{Baran}, C.~{Benoist}, M.~{Birkinshaw}, A.~{Bongiorno}, M.N. {Bremer} et~al., \aap\ \textbf{592}, A1 (2016)

\bibitem{Pacaud2016}
F.~{Pacaud}, N.~{Clerc}, P.A. {Giles}, C.~{Adami}, T.~{Sadibekova}, M.~{Pierre}, B.J. {Maughan}, M.~{Lieu}, J.P. {Le F{\`e}vre}, S.~{Alis} et~al., \aap\ \textbf{592}, A2 (2016)

\bibitem{Adami2018}
C.~{Adami}, P.~{Giles}, E.~{Koulouridis}, F.~{Pacaud}, C.A. {Caretta}, M.~{Pierre}, D.~{Eckert}, M.E. {Ramos-Ceja}, F.~{Gastaldello}, S.~{Fotopoulou} et~al., \aap\ \textbf{620}, A5 (2018)

\bibitem{Adam2015}
R.~{Adam}, B.~{Comis}, J.F. {Mac{\'\i}as-P{\'e}rez}, A.~{Adane}, P.~{Ade}, P.~{Andr{\'e}}, A.~{Beelen}, B.~{Belier}, A.~{Beno{\^\i}t}, A.~{Bideaud} et~al., \aap\ \textbf{576}, A12 (2015)

\bibitem{Adam2016}
R.~{Adam}, B.~{Comis}, I.~{Bartalucci}, A.~{Adane}, P.~{Ade}, P.~{Andr{\'e}}, M.~{Arnaud}, A.~{Beelen}, B.~{Belier}, A.~{Beno{\^\i}t} et~al., \aap\ \textbf{586}, A122 (2016)

\bibitem{Perotto2020}
L.~{Perotto}, N.~{Ponthieu}, J.F. {Mac{\'\i}as-P{\'e}rez}, R.~{Adam}, P.~{Ade}, P.~{Andr{\'e}}, A.~{Andrianasolo}, H.~{Aussel}, A.~{Beelen}, A.~{Beno{\^\i}t} et~al., \aap\ \textbf{637}, A71 (2020)

\bibitem{Gwyn2012}
S.D.J. {Gwyn}, \aj\ \textbf{143}, 38 (2012)

\bibitem{Aihara2022}
H.~{Aihara}, Y.~{AlSayyad}, M.~{Ando}, R.~{Armstrong}, J.~{Bosch}, E.~{Egami}, H.~{Furusawa}, J.~{Furusawa}, S.~{Harasawa}, Y.~{Harikane} et~al., \pasj\ \textbf{74}, 247 (2022)

\bibitem{Eckert2011}
D.~{Eckert}, S.~{Molendi}, S.~{Paltani}, \aap\ \textbf{526}, A79 (2011)

\bibitem{Eckert2020}
D.~{Eckert}, A.~{Finoguenov}, V.~{Ghirardini}, S.~{Grandis}, F.~{Kaefer}, J.~{Sanders}, M.~{Ramos-Ceja}, The Open Journal of Astrophysics \textbf{3}, 12 (2020)

\bibitem{Nagai2007}
D.~{Nagai}, A.V. {Kravtsov}, A.~{Vikhlinin}, \apj\ \textbf{668}, 1 (2007)

\bibitem{Navarro1996}
J.F. {Navarro}, C.S. {Frenk}, S.D.M. {White}, \apj\ \textbf{462}, 563 (1996)

\bibitem{PlanckXX2014}
{Planck Collaboration}, P.A.R. {Ade}, N.~{Aghanim}, C.~{Armitage-Caplan}, M.~{Arnaud}, M.~{Ashdown}, F.~{Atrio-Barandela}, J.~{Aumont}, C.~{Baccigalupi}, A.J. {Banday} et~al., \aap\ \textbf{571}, A20 (2014)

\end{thebibliography}

\end{document}